\documentclass[openacc]{rstransa}

\usepackage[square,numbers,sort&compress]{natbib}

\newcommand{\sis}{{Si~X/S~X}}
\newcommand{\caar}{{Ca~XIV/Ar~XIV}}
\newcommand{\fes}{{Fe~XVI/S~XIII}}

\newcommand{\cref}[1]{\textcolor{black}{#1}} 
\newcommand{\ccref}[1]{\textcolor{black}{#1}} 

\titlehead{Research}

\begin{document}

\title{Comparing First Ionisation Potential bias diagnostics in the solar atmosphere}

\author{K.~D.~Spruksta$^{1}$, D.~M.~Long$^{1,2}$, A.~S.~H.~To$^{3}$}

\address{$^{1}$Centre for Astrophysics and Relativity, School of Physical Sciences, Dublin City University\\
$^{2}$Astronomy \& Astrophysics Section, Dublin Institute for Advanced Studies, Dublin D02 XF86, Ireland \\
$^{3}$European Space Agency (ESA), European Space Research and Technology Centre (ESTEC), Keplerlaan 1, 2201 AZ Noordwijk, The Netherlands}

\subject{astrophysics, solar physics}


\corres{David Long\\
\email{david.long@dcu.ie}}

\begin{abstract}
\cref{Plasma composition} in the solar atmosphere differs between the photosphere and corona, producing an observable difference in elemental abundance known as the FIP effect. The FIP effect is characterised by the ratio of low to high FIP elements, \cref{giving a number known as the FIP bias. FIP bias values vary between different regions of the solar atmosphere, with typical observed values of} $\sim$1 for coronal holes, $\sim$1.5-2 for the quiet Sun, and $\sim$3 for active regions. The Extreme ultraviolet Imaging Spectrometer (EIS) onboard the \emph{Hinode} spacecraft has enabled the widespread use of the \sis\ line pair as a FIP bias diagnostic, but EIS observes other line pairs that can be used to estimate FIP bias. We consider three FIP bias diagnostics observed by \emph{Hinode}/EIS (\sis, \caar, and \fes), comparing the FIP bias between \cref{Quiet} Sun and an \cref{Active} region. We also \cref{assume} a range of signal-to-noise (SNR) cutoff values for each pixel, finding that while the SNR cutoff affects the number of useable pixels, higher (lower) SNR cutoffs remove (retain) a tail of high FIP bias values within the measured distribution. However, the median value of the FIP bias distribution remains largely unchanged. These results show the importance of a more nuanced view of FIP bias when using this vitally important diagnostic rather than a simplistic one-size-fits-all approach.
\end{abstract}

\maketitle


\section{Introduction}
Analysis of spectroscopic observations of the solar atmosphere have revealed a distinct and measureable difference in elemental abundance between the solar photosphere and corona. This difference is particularly identifiable between the abundances of elements of high ($\geq$10~eV) and low ($<$10~eV) first ionisation potential (FIP), and is commonly referred to as the FIP bias \cite{Asplund:2009,Meyer:1985}. \cref{Once imprinted in a given region of the solar atmosphere the FIP bias generally remains conserved, although processes such as plasma mixing or gravitational settling can cause changes in plasma composition. Overall, this general conservation makes FIP bias a very useful diagnostic for identifying the origin of e.g., solar wind plasma \cite{Yardley:2024}, and has enabled the connection science promulgated by the ESA/NASA \emph{Solar Orbiter} spacecraft \cite{Mueller:2020}.}

The process by which plasma is fractionated in the solar atmosphere remains the subject of intense research, with the leading theory suggesting that waves in coronal loops induce a ponderomotive force at the loop footpoints in the interface region between the solar photosphere and corona. This ponderomotive force then acts on ions that are largely low-FIP, transporting them upward into the corona and leaving the largely neutral high-FIP elements behind, thus producing the observed FIP bias \cite[cf.][]{Laming:2015}. Previous observations have identified signatures in the photosphere of the waves predicted to induce a ponderomotive associated with coronal signatures of enhanced FIP bias \cite[e.g.,][]{Stangalini:2021,Murabito:2024,Baker:2021,Murabito:2021}. However, although there have been some indications of chromospheric and transition region signatures of related processes \cite[cf.][]{Long:2024,Testa:2023}, no direct observations have yet been made of the fractionation process.

Previous work, particularly using the spatially resolved observations provided by the Extreme ultraviolet Imaging Spectrometer \cite[EIS;][]{Culhane:2007} onboard the \emph{Hinode} spacecraft \cite{Kosugi:2007} has shown that different regions of the solar atmosphere have different FIP bias, with some correlation found between increasing magnetic field strength and FIP bias \cite{Brooks:2015,Baker:2018,Baker:2013,Mihailescu:2022,To:2023}. This is also consistent with historical observations using spectroheliograms from \emph{Skylab} \cite{Feldman:1993}. \cref{The FIP bias of the solar atmosphere is therefore a powerful plasma diagnostic that has been identified as a possible tracer of plasma evolution through the solar atmosphere \cite{Laming:2015}. As a result, the \emph{Solar Orbiter} spacecraft has been designed to include two instruments that can measure FIP bias using remote-sensing (the SPectral Imaging of the Coronal Environment, SPICE, spectrometer \cite{Spice:2020}), and \emph{in-situ} (the Solar Wind Analyser, SWA, particle detector \cite{Owen:2020}) measurements.} 

However, although all of these instruments measure FIP bias using a combination of high- and low-FIP elements, the combinations of elements are different in each case, sensitive to different temperatures and ionisation fraction in the chromosphere. \cref{Previous work has also tended to assume single values for FIP bias measurements of different regions of the solar atmosphere. The cores of active regions are widely assumed to have FIP bias values of $\sim$3 \cite{delzanna:2014}, while the quiet Sun has a typical FIP bias of $\sim$1.5-2 \cite{Ko:2016,Baker:2013} Ensuring that the widely used FIP bias diagnostics behave in a manner consistent with their temperature sensitivities in different regions of the solar atmosphere \cite{Brooks:2015, Baker:2018, Baker:2013, Mihailescu:2022, To:2023}, and provide interpretable FIP bias values of those different regions, is therefore vital for our understanding of this increasingly important parameter.}

Here we compare three commonly used FIP bias diagnostics from the \emph{Hinode}/EIS instrument, namely the \sis, \fes, and \caar\ line ratios, focussing on how they behave in distinct regions of the solar atmosphere. Using \cref{a} full-disk \cref{mosaic} taken between 18-20~October~2015, we identified a region of \cref{Quiet Sun} and an \cref{Active} region and compared the distribution of FIP bias values in those regions for each diagnostic studied. The observations and data are presented in Section~\ref{s:obs}, with the results outlined in Section~\ref{s:res}, before some conclusions are drawn in Section~\ref{s:conc}.

\section{Observations}\label{s:obs}

\begin{figure}
    \includegraphics[width=1\linewidth]{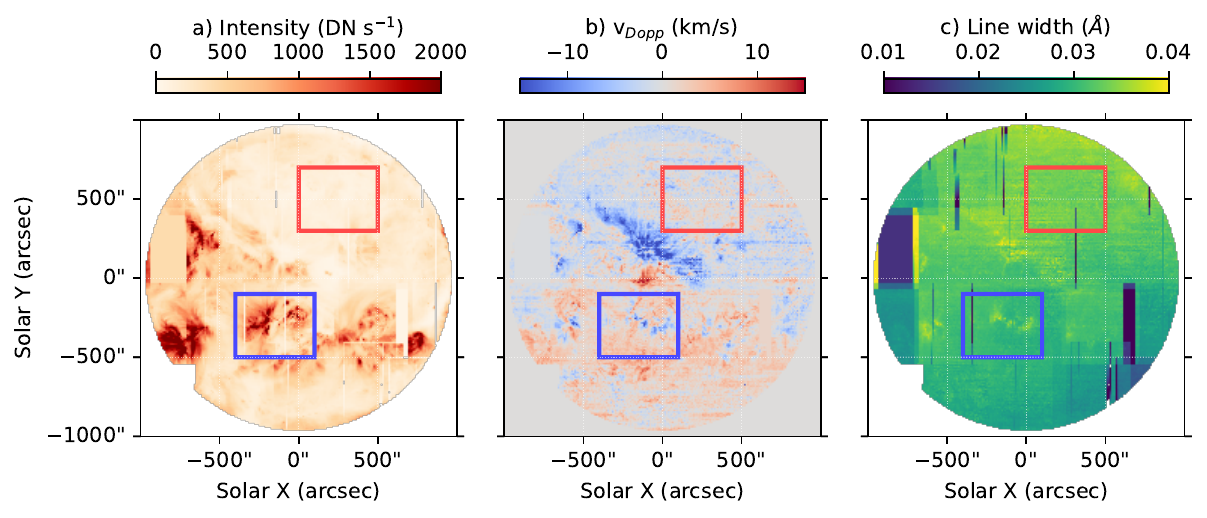}
    \caption{Full disk mosaics taken from 18-20~Oct~2015 by \emph{Hinode}/EIS showing the fitted peak intensity (panel~a), Doppler velocity (panel~b), and line width (panel~c) of the Fe~XII~195.119~\AA\ emission line. The \cref{red} (blue) box shows the \cref{Quiet} Sun (active region) region of interest further examined in Section~\ref{s:res}.}
    \label{fig:context}
\end{figure}

The set of raster scans comprising the full disk observations described here were taken by the Extreme ultraviolet Imaging Spectrometer \citep[EIS;][]{Culhane:2007} onboard the \emph{Hinode} spacecraft \citep{Kosugi:2007} from 18-20 October 2015. \emph{Hinode}/EIS regularly takes full-disk mosaic scans of the solar corona across a range of Extreme UltraViolet (EUV) lines, enabling multiple research opportunities including a comparison to be made between active regions of different ages \citep[cf.][]{Mihailescu:2022}, and the first full disk FIP maps \citep{Brooks:2015}. Figure~\ref{fig:context} shows the peak fitted intensity (left panel), Doppler velocity (middle panel) and nonthermal velocity (right panel) of the Fe~XII~195.119~\AA\ emission line for this full disk raster mosaic. It is clear from Figure~\ref{fig:context} that the solar disk at the time included a number of active regions and a large region of quiet Sun, enabling a direct comparison to be made between the different \cref{FIP} bias diagnostics measured here in these different regions. 


The full-disk mosaics used here were produced using a series of 26 pointings of the \emph{Hinode} spacecraft taken between 10:27:19~UT on 18-October-2015 and 00:26:12~UT on 20-October-2015, in each case using the DHB\_007 EIS study. This produces 123 pointings per raster, with a 4" step using the 2" slit, taking observations in 25 spectral windows. Each of the individual EIS rasters were processed and fitted using the eispac python package \cite{Weberg:2023}, with the fitted spectral lines then used to calculate the elemental abundance using several different diagnostics. The most commonly used FIP bias diagnostic produced by this full disk spectral scan is the low-FIP Si~X~258.37~\AA/\cref{intermediate}-FIP S~X~264.23~\AA\ diagnostic previously used by \cite{Brooks:2015} to probe the origin of the slow solar wind. However, this study also contains the spectral lines required to produce the low-FIP Ca~XIV~193.87~\AA/high-FIP Ar~XIV~194.40~\AA\ and the low-FIP Fe~XVI~262.98~\AA/\cref{intermediate}-FIP S~XIII~256.69~\AA\ FIP bias diagnostics \cite{Feldman:2009}. 

\begin{figure}
    \includegraphics[width=1\linewidth]{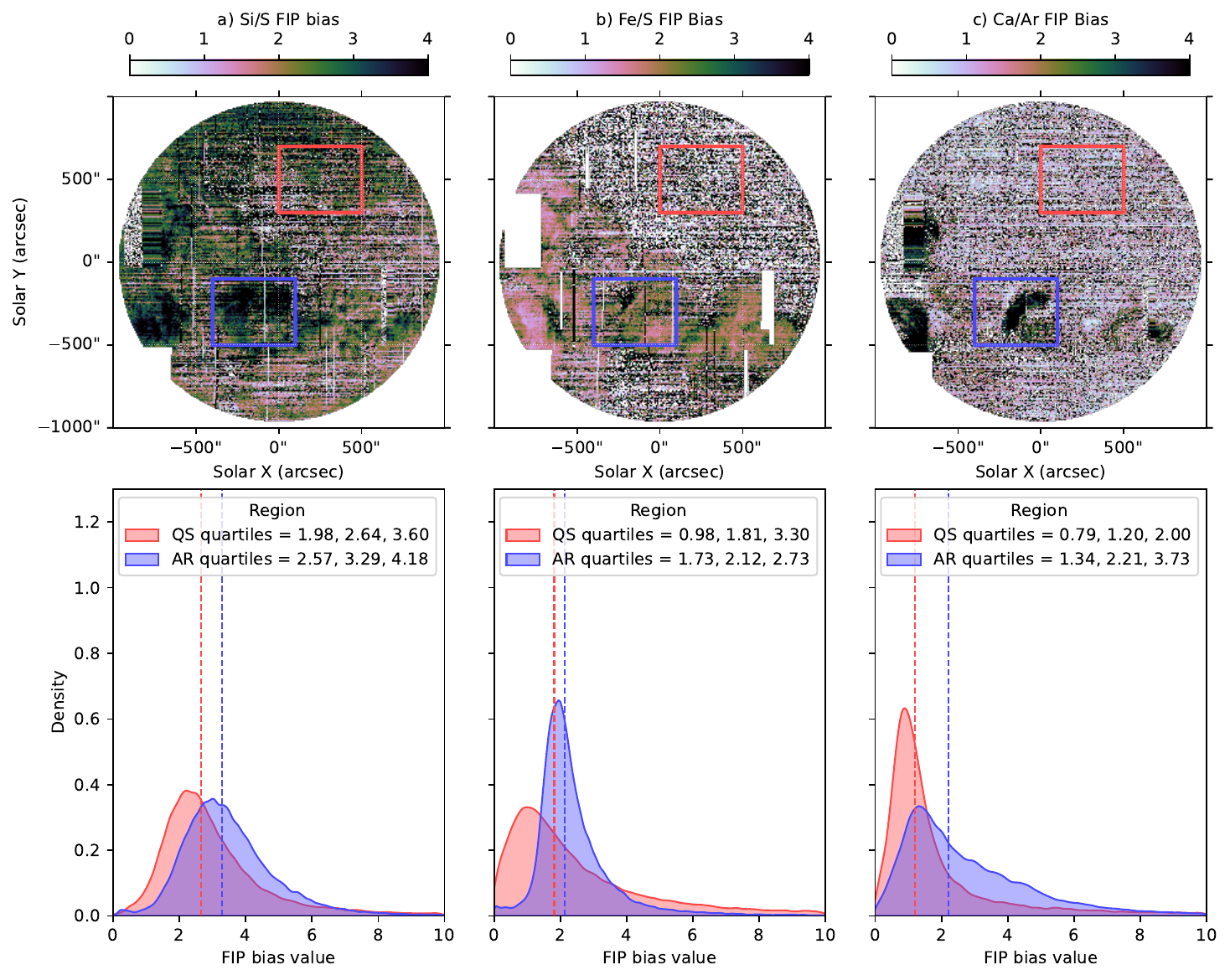}
    \caption{Top row: Full disk mosaics taken by \emph{Hinode}/EIS showing the derived \sis\ (panel~a, left), \fes\ (panel~b, centre), and \caar\ (panel~c, right) FIP maps. The \cref{red} (blue) box shows the \cref{Quiet} Sun (active region) region of interest. Bottom row shows the corresponding Kernel Density Estimator (KDE) plots of the distributions corresponding to the \cref{Quiet} Sun (\cref{red}) and active region (blue) from the derived \sis\ (left), \fes\ (centre), and \caar\ (right) FIP maps. \cref{The legend gives the 25$^{th}$, 50$^{th}$ (i.e., median), and 75$^{th}$ percentile values for each distribution}}
    \label{fig:composition}
\end{figure}

We used the Markov Chain Monte Carlo technique developed by \cite{Brooks:2015} and previously outlined in detail by \cite{To:2021,Mihailescu:2022} to process the individual EIS rasters and produce the full-disk \sis\ FIP bias map\cref{\footnote{Code available at \url{https://github.com/andyto1234/demcmc_FIP}}}. This approach fits spectral lines from consecutive ionisation stages of Fe~VIII - Fe~XVII using either single or multiple Gaussians as appropriate. Note that typically single Gaussians are used unless the line is blended. These lines have formation temperatures in the range $\sim$0.5--5.5~MK, and the combination of fitted line intensities can then be used to calculate the differential emission measure, and estimate the FIP bias assuming \cref{an electron} density estimated using the Fe~XIII 202.04~\AA/203.83~\AA\ line ratio. This diagnostic is most sensitive to plasma with a temperature of T~$\sim$1-2~MK, corresponding to quiet Sun and \cref{quiescent active regions}. Note that this analysis was done using version~10 of the CHIANTI database \cite{Dere:1997,delzanna:2021} assuming the photospheric abundances of \cite{Scott:2015a,Scott:2015b} and \emph{Hinode}/EIS calibration of \cite{Warren:2014}. The result of this derivation is shown in panel~a of Figure~\ref{fig:composition}. It is clear from Figure~\ref{fig:composition}a that the active regions have higher FIP bias values of $\sim$4.

The second diagnostic, \caar, was used to probe slightly higher temperature plasma. The Ca~XIV and Ar~XIV lines have a higher formation temperature of $\sim$3.5~MK, corresponding to the hot cores of active regions. In contrast to the \cref{full DEM approach used to derive the} \sis\ diagnostic, the \caar\ diagnostic was derived by first fitting the individual spectral lines and then taking their ratio \cref{following the example of \cite{Baker:2019} who note that variations in the elemental abundances of the two elements primarily affect the magnitude of their contribution functions rather than their shape as a function of temperature. As a result, the intensity ratio can be used to derive the relative abundance between Ca and Ar \cite[see also][for further discussions]{Doschek:2015,Doschek:2016,Doschek:2017}.} For this diagnostic, shown in Figure~\ref{fig:composition}\cref{c}, a clear distinction can be made between the hotter active regions and the cooler quiet Sun. The hotter active regions display a higher FIP bias value of $\sim$4, with the cooler quiet Sun regions exhibiting a lower FIP bias value of $\le$1\cref{, with the image shown in Figure~\ref{fig:composition}c indicating that a lot of the pixels have FIP bias values of 0 that are ignored in this analysis}.

The final diagnostic examined here was the \fes\ diagnostic described and previously used by \cite{Baker:2022,To:2024}. The Fe~XVI and S~XIII lines were first fitted using single Gaussians as they are unblended, with the FIP bias calculated using the MCMC approach also used to estimate the \sis\ diagnostic. This process is described in more detail in \cite{Baker:2022,To:2024}. The Fe~XVI and S~XIII lines are sensitive to \cref{plasma} at T$\sim$2-3~MK, with the resulting full disk mosaic shown in Figure~\ref{fig:composition}\cref{b}. \cref{Figure~\ref{fig:composition}b shows stronger signal in the different active regions, with the quiet Sun regions showing a lot more white pixels. This indicates} that the \fes\ diagnostic is sensitive to active region and surrounding plasma, with a FIP bias of $\sim$2-3 observed for the different active regions in contrast to the FIP bias of $\le$1 seen for quiet Sun regions\cref{, again indicating that a lot of the pixels have FIP bias values of 0 that are ignored in this analysis (as for the \caar\ diagnostic above)}.

\section{Results}\label{s:res}

\subsection{Comparing different regions of the solar atmosphere}\label{ss:comparison}

\cref{To further investigate the different FIP bias diagnostics and better understand the importance of temperature and density sensitivity on the distribution of FIP bias values we defined two regions of interest for a more detailed examination. The red boxes in Figures~\ref{fig:context} and \ref{fig:composition} highlight a region of Quiet Sun, while the blue boxes highlight an active region. We then used a kernel density estimation (KDE) approach to compare the distribution of FIP bias values in the different regions \cite[cf.][]{Silverman:1986,Dacie:2016,Long:2024}. This approach does not require the use of user-defined bins \cite{dejager:1986}, and enables a clear identification of the different feature of the individual datasets.} It is clear from each of the three panels in \cref{the bottom row of} Figure~\ref{fig:composition} that as expected there is a difference in FIP bias between the regions defined as \cref{Quiet} Sun and Active Region. For each diagnostic studied, the distribution corresponding to the Quiet Sun (shown in \cref{red}) tends to peak at a lower FIP bias value than the distribution corresponding to the Active Region (shown in blue). This is to be expected, with the \cref{quiet} Sun typically expected to exhibit a FIP bias value of $\sim$1.5-2, corresponding to \cref{weakly fractionated} photospheric plasma \cite{Baker:2013}. However, there is a clear distribution of values for the \cref{Quiet} Sun region as measured using each diagnostic, with a strong tail of values having a value greater than 2 for each diagnostic, but particularly both the \sis\ and \fes\ diagnostics. The \cref{quartile} values of the distributions are also given in the legends, with \cref{Quiet} Sun median \cref{(i.e., 50$^{th}$ percentile)} values of \cref{2.64} (\sis), \cref{1.20} (\caar), and \cref{1.81} (\fes). 

While active regions are considered to have higher FIP bias values of $\sim$3 \cite{delzanna:2014}, the distributions shown in Figure~\ref{fig:composition} are quite broad, with tails extending to FIP bias values of $\sim$\cref{10}. It is interesting that the \cref{25$^{th}$ to 75$^{th}$ percentiles of the} \sis\ distribution \cref{(containing 50\% of the values of the distribution) extend from $\sim$2.57-4.18}, with a median value of \cref{3.29}. The \caar\ diagnostic also shows a distribution of FIP bias values in the active region, \cref{with the interquartile range of the distribution} extending from \cref{1.34-3.73}, with a tail of values reaching $\sim$\cref{10}. The median FIP bias value here is at \cref{2.21}, much lower than that expected for active regions, which is surprising given that the \caar\ diagnostic is much more sensitive to the hotter plasma expected in active regions of the solar atmosphere. The distribution of \cref{the bulk of the} FIP bias values for the \fes\ diagnostic is much narrower, \cref{with the interquartile range} extending from \cref{1.73-2.73}, but with a median value of \cref{2.12} comparable to that of the corresponding \cref{Quiet} Sun region. 

\subsection{Examining the signal-to-noise ratio}\label{ss:choice}

\begin{figure}
    \includegraphics[width=1\linewidth]{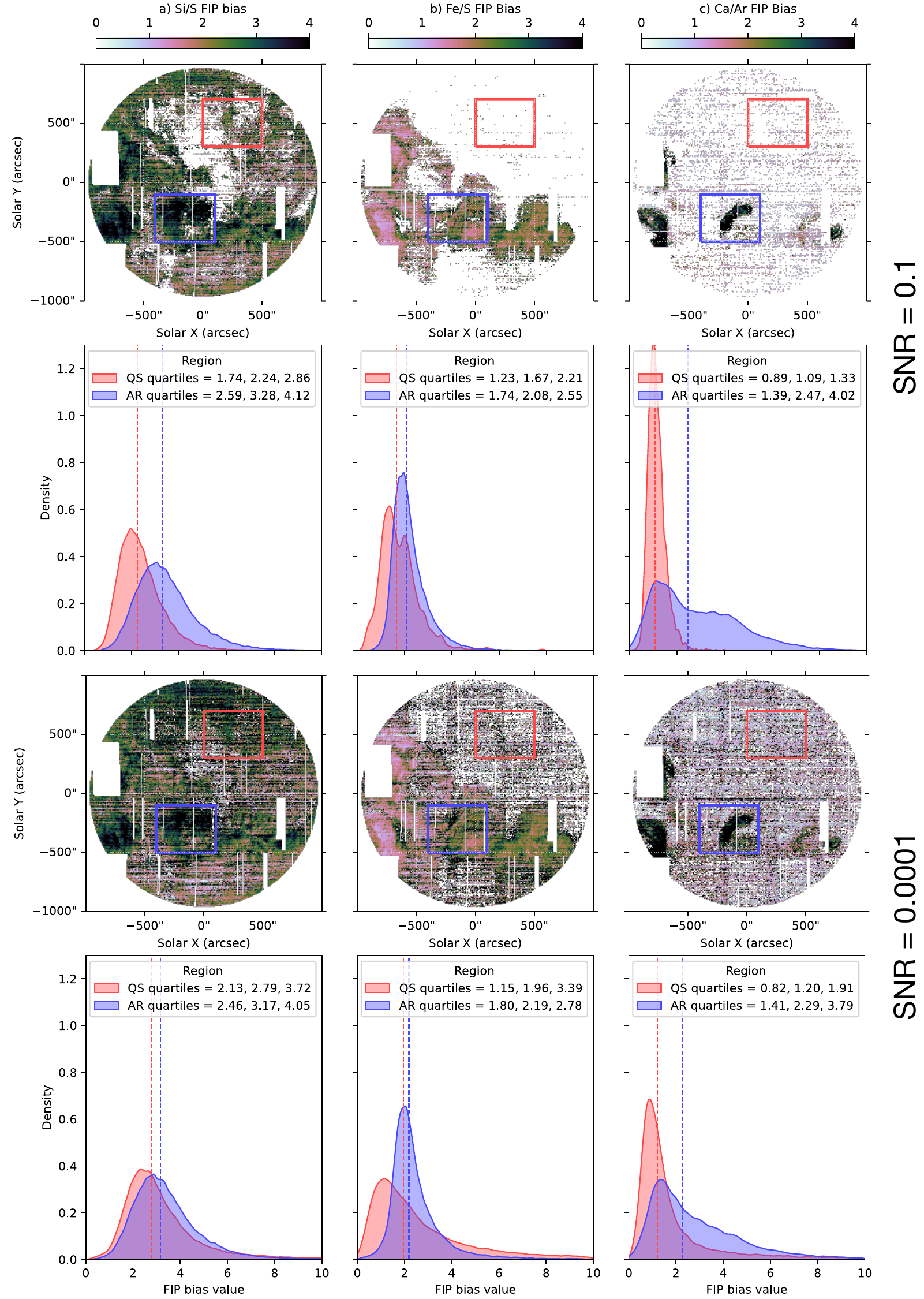}
    \caption{\emph{Hinode}/EIS full disk mosaics and corresponding KDE plots for signal-to-noise cutoffs of 0.1 (upper two rows) and 0.0001 (bottom two rows). In each case, the FIP bias ratios are (from left to right) \sis, \fes, and \caar.}
    \label{fig:comp_SNR_0}
\end{figure}

Although the images shown in the top row of Figure~\ref{fig:composition} suggest that the three FIP bias ratios measured provide \cref{distinct measurements for the different regions} across the solar disk, \cref{the \fes\ and \caar\ ratios are high-temperature emission lines that should produce little to no measurable signal in quiet Sun regions and coronal holes in particular} \ccref{\cite[cf.][]{Young:2022}}. To investigate this further, we examined the different FIP bias ratios using a filtering method to remove pixels with a signal-to-noise ratio below a range of pre-defined levels. This approach ensures that only pixels with a sufficiently strong signal in the different lines used to estimate the FIP bias ratios were considered in each case. It also enables a better understanding of the distribution of FIP bias values in the regions of interest examined in this work. 

The signal-to-noise \cref{ratio} (SNR) for each pixel was estimated by first subtracting the fit \cref{to the intensity} for the individual spectral lines from the observed \cref{intensity} to calculate the residuals. The SNR was then taken as the \cref{spectral intensity divided by the} standard deviation of the residuals \cref{$\sigma_{res}$ as so;
\begin{equation}
    SNR = \frac{I - fit}{\sigma_{res}}.
\end{equation}}
We then defined a range of cutoff values, in each case only including those pixels with an SNR above the cutoff \cref{for} each wavelength. Figure~\ref{fig:comp_SNR_0} shows the corresponding images and KDE plots of the regions of interest for the different FIP bias ratios studied assuming a SNR cutoff of 0.1 (upper two rows) and 0.0001 (bottom two rows). 

\begin{figure}
    \includegraphics[width=1\linewidth]{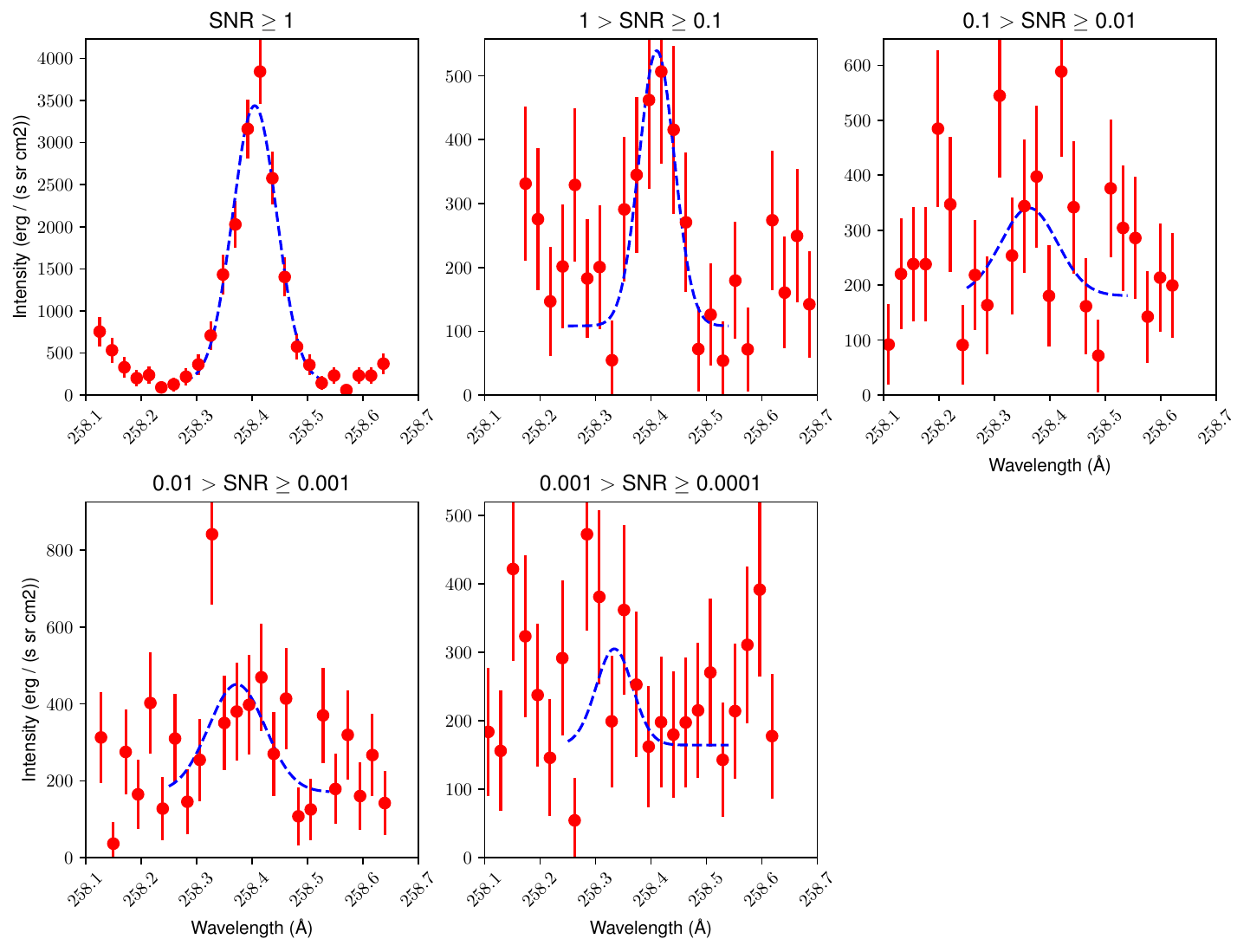}
    \caption{\cref{A comparison of fits to the Si~X~258.37~\AA\ spectral line for different ranges of the signal-to-noise (in each case given in the panel title).}}
    \label{fig:SNR_spectral}
\end{figure}

\ccref{To fully examine the importance of filtering FIP bias estimates by comparing signal-to-noise values,} a range of SNR cutoff values \ccref{from high to very low values that approximated noise} were examined as part of this work. A SNR cutoff of 5 meant that no valid data was returned for any wavelength considered here, whereas a SNR cutoff of 1 only revealed a small number of useable pixels scattered throughout the \sis\ and \fes\ FIP bias images. Using a SNR cutoff of 0.1 (as shown in Figure~\ref{fig:comp_SNR_0}) provides a much larger number of useable pixels, particularly in the \sis\ image and distribution. The white area seen in the top left panel of Figure~\ref{fig:comp_SNR_0} corresponds to a coronal hole, with a correspondingly low observable signal. For the \fes\ and particularly the \caar\ FIP bias ratios, there is very little signal outside the active regions, with very narrow associated distributions of values for the quiet Sun regions in the corresponding KDE plots. It is interesting that the KDE for the \caar\ active region appears to have two peaks, with the image showing a very strong signal in the core of the active region and a weaker signal in the outer part of the active region. \cref{Figure~\ref{fig:SNR_spectral} shows the fit to the Si~X~258.37~\AA\ spectral line for different ranges of the signal-to-noise ratio studied here. As expected, the fit to the spectral line for SNR>1 is quite good, with the fits to the spectral line becoming progressively worse for lower SNR ranges.}

Decreasing the SNR cutoff to 0.0001 (bottom two rows of Figure~\ref{fig:comp_SNR_0}) produces a set of FIP bias images much closer to that seen in Figure~\ref{fig:composition}\ccref{, for which no SNR filtering has been applied}. The \sis\ image has not changed much, albeit with the inclusion of more signal in the coronal hole region. The associated KDE plots show that the distributions of values from the \cref{Quiet} Sun and active region are much closer together with the lower SNR cutoff, although the median values of the distributions are still comparable. For the \fes\ FIP bias, there are now many more useable pixels across the quiet Sun, with a much smoother distribution of values in the KDE plots for both \cref{Quiet} Sun and active region. However, the median values of both distributions remain comparable. Finally, the \caar\ image again shows many more useable pixels across the solar disk, with the previously twin peaked active region KDE plot now smoothed out to produce a more gradually decreasing slope towards higher FIP values. Again, the median FIP bias values are comparable in each case. 

It is interesting that the KDE distributions for the lower SNR cutoff have much longer tails of higher FIP bias values, indicating that more noisy pixels producing spurious FIP bias values are now being counted and included in the distributions. However, the choice of median rather than mean to estimate a single value for the FIP bias negates this trend, giving estimated single FIP bias values that are effectively unchanged for different SNR cutoff values.

\section{Discussion \& Conclusions}\label{s:conc}

As mass and energy evolve through the solar atmosphere into the heliosphere, the properties of the plasma change and evolve as the result of turbulence, heating, mass motion, and elemental fractionation. While temperature, density, and other plasma properties can change as the plasma propagates out into the heliosphere, the FIP bias of the plasma remains fixed, imprinted by the properties of the region where the plasma was initially fractionated. This makes FIP bias a valuable diagnostic tool for tracking and quantifying the evolution of plasma from the Sun into the heliosphere, enabling a connection to be made between remote-sensing observations of the Sun and direct measurements of the solar wind \emph{in-situ} at spacecraft throughout the heliosphere. The \emph{Solar Orbiter} spacecraft in particular has been designed to leverage this plasma property and thus maximise the scientific return, while it also has important implications for the forecasting and understanding of space weather and the effect of the Sun on the near-Earth environment.

FIP bias measurements of different parts of the solar atmosphere are typically given as single values, with active regions, quiet Sun, and coronal holes typically quoted as having FIP bias values of $\sim$3, $\sim$1.5-2 and $\sim$1 respectively \cite{Brooks:2015}. This can be interpreted as active regions having coronal composition, coronal holes having photospheric composition, and the quiet Sun \cref{having a composition that lies between coronal and photospheric values.} However FIP bias does not have a single unique measurement, as any combination of low and high FIP elements can be used to estimate FIP bias \cite[accounting for comparable emissivities, temperature, and density effects, e.g.,][]{Feldman:2009}. As a result, there are multiple potential line pairs within the spectral window observed by the \emph{Hinode}/EIS instrument that can and have been used to observe the solar corona. Although the \sis\ diagnostic is the most commonly used, it is also possible to use the \caar\ and \fes\ line pairs to estimate the FIP bias, with each line pair sensitive to plasma at slightly different temperatures. 

\ccref{Previous work \cite{Mihailescu:2022} has suggested that FIP bias may be best observed as a distribution rather than the commonly used single value approach.} This has implications for \cref{how it can be used to relate plasma observed on the Sun with plasma detected \emph{in-situ} (e.g., as by the \emph{Solar Orbiter} spacecraft)}, and as a result requires further investigation. In this work we used full disk mosaic spectral scans of the Sun to compare FIP bias measurements using three different diagnostics; \sis, \caar, and \fes. As shown in Figure~\ref{fig:composition}, \cref{the full disk spectral scans highlight the different regions of the solar disk, because each spectral line has different temperature sensitivities, allowing the active region and quiet sun to be distinguished. While each diagnostic was applied to the full disk, as shown in Figure~\ref{fig:composition} t}he \sis\ diagnostic, sensitive to coronal plasma of $\sim$1-2~MK, has signal and measureable FIP bias in both the \cref{Quiet} Sun and active regions, whereas the \caar\ diagnostic, sensitive to hot plasma of $\sim$3-4~MK exhibits a clear FIP bias \cref{primarily} in the active regions. The \fes\ diagnostic is sensitive to plasma between these two extremes, identifying the full \cref{extent} of the active regions rather than just the cores seen using the \caar\ diagnostic.

However, a different picture emerges when considering the distribution of values within regions of interest defined to compare the quiet Sun and active regions. As shown in Figure~\ref{fig:composition}, the distributions of FIP bias values measured in the quiet Sun \cref{tends} to peak at a slightly lower value than \cref{the distribution in} the active region. It is notable that the bulk of the FIP bias values in the active region distributions are lower than three for \cref{the \caar\ and \fes\ diagnostics}. The choice of elements and lines used to estimate the FIP bias is also interesting and worth considering when trying to compare the different regions of interest. The three diagnostics examined here were chosen as they are well observed by \emph{Hinode}/EIS and are therefore commonly used to produce spatially resolved FIP bias maps of the solar atmosphere. \cref{However, the results described here show that there are distinct variations between the individual diagnostics which must be acknowledged when comparing them, with each diagnostic affected differently by variations in plasma properties such as temperature and density. In addition, the three diagnostics are derived in slightly different ways, with a DEM technique used to derive the \sis\ and \fes\ diagnostics while the \caar\ diagnostic is typically derived using an intensity ratio approach due to the comparable contribution functions of the two spectral lines. This could also contribute to the observed differences, although \cite{Baker:2019} previously found that the intensity ratio approach was comparable to the DEM approach for the \caar\ diagnostic. Work elsewhere in this special issue has also highlighted important characteristics of the plasma to consider when comparing different diagnostics. In particular,} \cite{Orlovskij:2025} \cref{suggest} that the behaviour of Sulphur is sensitive to the ambient magnetic field strength, while work by \cite{Reep:2025} also in this special issue \cref{shows} the importance of considering the temporal evolution of ionisation fraction when working with elemental abundances. 

In addition, \cref{the visual representation of the derived FIP bias can be} affected by the assumed signal-to-noise ratio used to discount noisy pixels or pixels with no useable signals. \cref{The DEM technique commonly used to derive the \sis\ and \fes\ diagnostics discounts pixels where the Markov Chain Monte Carlo analysis has a $\chi^2$ value greater than the number of lines used for the calculation \cite[cf.][]{Brooks:2011,Brooks:2015}. However, filtering of data assuming a defined signal-to-noise ratio is not typically performed.} Figure~\ref{fig:comp_SNR_0} compares the different FIP bias ratios for varying SNR values of 0.1 and 0.0001, finding that while the different SNR cutoff values clearly affect the \cref{the FIP bias maps and the} number of useable pixels in each case, the corresponding KDE distributions are largely unaffected. A lower SNR cutoff contribute\cref{s} more noisy pixels with higher FIP bias values to the KDE distributions, producing a longer tail on the distributions. However, despite the inclusion of more high FIP bias values, the median values of the distributions exhibit\cref{s} very little change. \cref{A comparison of the fits to the Si~X~258.37~\AA\ spectral line for different ranges of the signal-to-noise is shown in Figure~\ref{fig:SNR_spectral}. It is clear that the quality of fits to the spectral lines decreases with lower SNR ranges, underlining the importance of checking individual fits and ensuring a robust approach to including pixels with lower SNR when undertaking this analysis. Note that the SNR cutoff values yielding ``useable'' measurements here are much lower than those used in other contexts. The aim here is to highlight the importance of understanding the data and its limitations, and ensuring the correct use of appropriate diagnostics when comparing different regions of the solar atmosphere.} \ccref{It is also worth noting that previous work by \cite{Young:2022} found negligible emission from quiet Sun and coronal holes in higher temperature spectral lines observed by EIS. Instead, they suggested that corresponding signal in these lines must be scattered light from nearby active regions.}

These results indicate the importance of examining fully all aspects of the data when using FIP bias ratios. It is misleading to quote a fixed FIP bias value rather than a distribution when considering plasma fractionation and to quote fixed FIP bias values for defined regions of the solar atmosphere based on a single line pair diagnostic. \cref{Instead, this work shows that quoting the quartiles of the FIP bias distribution for a given region of interest may be more useful, as this makes no assumptions of the shape of the distribution, provides an indication of its width and asymmetry, and enables a more statistically rigorous analysis of the distributions.} \ccref{Previous work by \cite{Brooks:2015} did derive an uncertainty of $\sim$0.3 for FIP bias measurements given the uncertainy in EIS intensity measurements and the nature of the MCMC DEM approach that they used. However, the distributions of FIP bias values presented here show significantly more variation, suggesting that while the uncertainty derived by \cite{Brooks:2015} may be correct for an individual pixel, the quartile approach may provide a more holistic estimate of the uncertainty within a region of interest.} The signal-to-noise of the data must also be considered, \cref{as the work discussed here shows that it offers an additional check on the validity of the FIP bias diagnostic, particularly in regions with a temperature outside the optimal temperature range of the diagnostic being used.} 

As our understanding of the processes by which plasma is fractionated and transported through the solar atmosphere improve and evolve, it will be possible to better characterise the FIP bias of the different regions of the solar corona. The improved spectral, spatial, and temporal resolution of next generation instrumentation such as Solar-C EUVST and the ongoing work linking the Sun to the heliosphere with \emph{Solar Orbiter} \cite{Yardley:2024} will help in that regard.

\dataccess{No new data were generated as part of this study. The Hinode EIS data can be downloaded from the Hinode Science Data Centre (\href{http://sdc.uio.no/sdc/}{http://sdc.uio.no/sdc/})}

\aucontribute{DML conceived the study. KDS carried out the data reduction and scientific analysis with assistance from DML. ASHT processed the \fes\ data. DML drafted the manuscript with help from KDS. All authors read and approved the manuscript.}

\competing{The author(s) declare that they have no competing interests.}

\funding{}

\ack{
\cref{The authors wish to thank the referee whose suggestions helped to improve the paper.}
We thank the Royal Society for its support throughout the organization of the Theo Murphy meeting on "Solar Abundances in Space and Time" and production of this special issue. ASHT acknowledge support through the
European Space Agency (ESA) Research Fellowship Programme in Space Science. Hinode is a Japanese mission developed and launched by ISAS/JAXA, collaborating with NAOJ as a domestic partner, and NASA and STFC (UK) as international partners. Scientific operation of Hinode is performed by the Hinode science team organized at ISAS/JAXA. Support for the post-launch operation is provided by JAXA and NAOJ (Japan), STFC (UK), NASA, ESA, and NSC (Norway).
This research used the following software packages during analysis of the data and preparation of this paper: Numpy \citep{Harris:2020}, Scipy \cite{Virtanen:2020}, SunPy \citep{Sunpy:2020}, Seaborn \citep{Waskom:2021}, Matplotlib \citep{Hunter:2007}, Astropy \citep{Astropy:2022}, eispac \citep{Weberg:2023}.
}


\bibliographystyle{RS} 
\bibliography{bibliography} 

@ARTICLE{Asplund:2009,
       author = {{Asplund}, Martin and {Grevesse}, Nicolas and {Sauval}, A. Jacques and {Scott}, Pat},
        title = "{The Chemical Composition of the Sun}",
      journal = {\araa},
         year = 2009,
        month = sep,
       volume = {47},
       number = {1},
        pages = {481-522},
          doi = {10.1146/annurev.astro.46.060407.145222},
archivePrefix = {arXiv},
       eprint = {0909.0948},
 primaryClass = {astro-ph.SR},
       adsurl = {https://ui.adsabs.harvard.edu/abs/2009ARA&A..47..481A}
}

@ARTICLE{Astropy:2022,
       author = {{Astropy Collaboration} and {Price-Whelan}, Adrian M. and {Lim}, Pey Lian and {Earl}, Nicholas and {Starkman}, Nathaniel and {Bradley}, Larry and {Shupe}, David L. and {Patil}, Aarya A. and {Corrales}, Lia and {Brasseur}, C.~E. and {N{\"o}the}, Maximilian and {Donath}, Axel and {Tollerud}, Erik and {Morris}, Brett M. and {Ginsburg}, Adam and {Vaher}, Eero and {Weaver}, Benjamin A. and {Tocknell}, James and {Jamieson}, William and {van Kerkwijk}, Marten H. and {Robitaille}, Thomas P. and {Merry}, Bruce and {Bachetti}, Matteo and {G{\"u}nther}, H. Moritz and {Aldcroft}, Thomas L. and {Alvarado-Montes}, Jaime A. and {Archibald}, Anne M. and {B{\'o}di}, Attila and {Bapat}, Shreyas and {Barentsen}, Geert and {Baz{\'a}n}, Juanjo and {Biswas}, Manish and {Boquien}, M{\'e}d{\'e}ric and {Burke}, D.~J. and {Cara}, Daria and {Cara}, Mihai and {Conroy}, Kyle E. and {Conseil}, Simon and {Craig}, Matthew W. and {Cross}, Robert M. and {Cruz}, Kelle L. and {D'Eugenio}, Francesco and {Dencheva}, Nadia and {Devillepoix}, Hadrien A.~R. and {Dietrich}, J{\"o}rg P. and {Eigenbrot}, Arthur Davis and {Erben}, Thomas and {Ferreira}, Leonardo and {Foreman-Mackey}, Daniel and {Fox}, Ryan and {Freij}, Nabil and {Garg}, Suyog and {Geda}, Robel and {Glattly}, Lauren and {Gondhalekar}, Yash and {Gordon}, Karl D. and {Grant}, David and {Greenfield}, Perry and {Groener}, Austen M. and {Guest}, Steve and {Gurovich}, Sebastian and {Handberg}, Rasmus and {Hart}, Akeem and {Hatfield-Dodds}, Zac and {Homeier}, Derek and {Hosseinzadeh}, Griffin and {Jenness}, Tim and {Jones}, Craig K. and {Joseph}, Prajwel and {Kalmbach}, J. Bryce and {Karamehmetoglu}, Emir and {Ka{\l}uszy{\'n}ski}, Miko{\l}aj and {Kelley}, Michael S.~P. and {Kern}, Nicholas and {Kerzendorf}, Wolfgang E. and {Koch}, Eric W. and {Kulumani}, Shankar and {Lee}, Antony and {Ly}, Chun and {Ma}, Zhiyuan and {MacBride}, Conor and {Maljaars}, Jakob M. and {Muna}, Demitri and {Murphy}, N.~A. and {Norman}, Henrik and {O'Steen}, Richard and {Oman}, Kyle A. and {Pacifici}, Camilla and {Pascual}, Sergio and {Pascual-Granado}, J. and {Patil}, Rohit R. and {Perren}, Gabriel I. and {Pickering}, Timothy E. and {Rastogi}, Tanuj and {Roulston}, Benjamin R. and {Ryan}, Daniel F. and {Rykoff}, Eli S. and {Sabater}, Jose and {Sakurikar}, Parikshit and {Salgado}, Jes{\'u}s and {Sanghi}, Aniket and {Saunders}, Nicholas and {Savchenko}, Volodymyr and {Schwardt}, Ludwig and {Seifert-Eckert}, Michael and {Shih}, Albert Y. and {Jain}, Anany Shrey and {Shukla}, Gyanendra and {Sick}, Jonathan and {Simpson}, Chris and {Singanamalla}, Sudheesh and {Singer}, Leo P. and {Singhal}, Jaladh and {Sinha}, Manodeep and {Sip{\H{o}}cz}, Brigitta M. and {Spitler}, Lee R. and {Stansby}, David and {Streicher}, Ole and {{\v{S}}umak}, Jani and {Swinbank}, John D. and {Taranu}, Dan S. and {Tewary}, Nikita and {Tremblay}, Grant R. and {de Val-Borro}, Miguel and {Van Kooten}, Samuel J. and {Vasovi{\'c}}, Zlatan and {Verma}, Shresth and {de Miranda Cardoso}, Jos{\'e} Vin{\'\i}cius and {Williams}, Peter K.~G. and {Wilson}, Tom J. and {Winkel}, Benjamin and {Wood-Vasey}, W.~M. and {Xue}, Rui and {Yoachim}, Peter and {Zhang}, Chen and {Zonca}, Andrea and {Astropy Project Contributors}},
        title = "{The Astropy Project: Sustaining and Growing a Community-oriented Open-source Project and the Latest Major Release (v5.0) of the Core Package}",
      journal = {\apj},
         year = 2022,
        month = aug,
       volume = {935},
       number = {2},
          eid = {167},
        pages = {167},
          doi = {10.3847/1538-4357/ac7c74},
archivePrefix = {arXiv},
       eprint = {2206.14220},
 primaryClass = {astro-ph.IM},
       adsurl = {https://ui.adsabs.harvard.edu/abs/2022ApJ...935..167A}
}

@ARTICLE{Baker:2013,
       author = {{Baker}, D. and {Brooks}, D.~H. and {D{\'e}moulin}, P. and {van Driel-Gesztelyi}, L. and {Green}, L.~M. and {Steed}, K. and {Carlyle}, J.},
        title = "{Plasma Composition in a Sigmoidal Anemone Active Region}",
      journal = {\apj},
         year = 2013,
        month = nov,
       volume = {778},
       number = {1},
          eid = {69},
        pages = {69},
          doi = {10.1088/0004-637X/778/1/69},
archivePrefix = {arXiv},
       eprint = {1310.0999},
 primaryClass = {astro-ph.SR},
       adsurl = {https://ui.adsabs.harvard.edu/abs/2013ApJ...778...69B}
}

@ARTICLE{Baker:2018,
       author = {{Baker}, Deborah and {Brooks}, David H. and {van Driel-Gesztelyi}, Lidia and {James}, Alexander W. and {D{\'e}moulin}, Pascal and {Long}, David M. and {Warren}, Harry P. and {Williams}, David R.},
        title = "{Coronal Elemental Abundances in Solar Emerging Flux Regions}",
      journal = {\apj},
         year = 2018,
        month = mar,
       volume = {856},
       number = {1},
          eid = {71},
        pages = {71},
          doi = {10.3847/1538-4357/aaadb0},
archivePrefix = {arXiv},
       eprint = {1801.08424},
 primaryClass = {astro-ph.SR},
       adsurl = {https://ui.adsabs.harvard.edu/abs/2018ApJ...856...71B}
}

@ARTICLE{Baker:2019,
       author = {{Baker}, Deborah and {van Driel-Gesztelyi}, Lidia and {Brooks}, David H. and {Valori}, Gherardo and {James}, Alexander W. and {Laming}, J. Martin and {Long}, David M. and {D{\'e}moulin}, Pascal and {Green}, Lucie M. and {Matthews}, Sarah A. and {Ol{\'a}h}, Katalin and {K{\H{o}}v{\'a}ri}, Zsolt},
        title = "{Transient Inverse-FIP Plasma Composition Evolution within a Solar Flare}",
      journal = {\apj},
         year = 2019,
        month = apr,
       volume = {875},
       number = {1},
          eid = {35},
        pages = {35},
          doi = {10.3847/1538-4357/ab07c1},
archivePrefix = {arXiv},
       eprint = {1902.06948},
 primaryClass = {astro-ph.SR},
       adsurl = {https://ui.adsabs.harvard.edu/abs/2019ApJ...875...35B}
}

@ARTICLE{Baker:2021,
       author = {{Baker}, Deborah and {Stangalini}, Marco and {Valori}, Gherardo and {Brooks}, David H. and {To}, Andy S.~H. and {van Driel-Gesztelyi}, Lidia and {D{\'e}moulin}, Pascal and {Stansby}, David and {Jess}, David B. and {Jafarzadeh}, Shahin},
        title = "{Alfv{\'e}nic Perturbations in a Sunspot Chromosphere Linked to Fractionated Plasma in the Corona}",
      journal = {\apj},
         year = 2021,
        month = jan,
       volume = {907},
       number = {1},
          eid = {16},
        pages = {16},
          doi = {10.3847/1538-4357/abcafd},
archivePrefix = {arXiv},
       eprint = {2012.04308},
 primaryClass = {astro-ph.SR},
       adsurl = {https://ui.adsabs.harvard.edu/abs/2021ApJ...907...16B}
}

@ARTICLE{Baker:2022,
       author = {{Baker}, D. and {Green}, L.~M. and {Brooks}, D.~H. and {D{\'e}moulin}, P. and {van Driel-Gesztelyi}, L. and {Mihailescu}, T. and {To}, A.~S.~H. and {Long}, D.~M. and {Yardley}, S.~L. and {Janvier}, M. and {Valori}, G.},
        title = "{Evolution of Plasma Composition in an Eruptive Flux Rope}",
      journal = {\apj},
         year = 2022,
        month = jan,
       volume = {924},
       number = {1},
          eid = {17},
        pages = {17},
          doi = {10.3847/1538-4357/ac32d2},
archivePrefix = {arXiv},
       eprint = {2110.11714},
 primaryClass = {astro-ph.SR},
       adsurl = {https://ui.adsabs.harvard.edu/abs/2022ApJ...924...17B}
}

@ARTICLE{Brooks:2011,
       author = {{Brooks}, David H. and {Warren}, Harry P.},
        title = "{Establishing a Connection Between Active Region Outflows and the Solar Wind: Abundance Measurements with EIS/Hinode}",
      journal = {\apjl},
         year = 2011,
        month = jan,
       volume = {727},
       number = {1},
          eid = {L13},
        pages = {L13},
          doi = {10.1088/2041-8205/727/1/L13},
archivePrefix = {arXiv},
       eprint = {1009.4291},
 primaryClass = {astro-ph.SR},
       adsurl = {https://ui.adsabs.harvard.edu/abs/2011ApJ...727L..13B}
}

@ARTICLE{Brooks:2015,
       author = {{Brooks}, David H. and {Ugarte-Urra}, Ignacio and {Warren}, Harry P.},
        title = "{Full-Sun observations for identifying the source of the slow solar wind}",
      journal = {Nature Communications},
         year = 2015,
        month = jan,
       volume = {6},
          eid = {5947},
        pages = {5947},
          doi = {10.1038/ncomms6947},
archivePrefix = {arXiv},
       eprint = {1605.09514},
 primaryClass = {astro-ph.SR},
       adsurl = {https://ui.adsabs.harvard.edu/abs/2015NatCo...6.5947B}
}

@ARTICLE{Culhane:2007,
       author = {{Culhane}, J.~L. and {Harra}, L.~K. and {James}, A.~M. and {Al-Janabi}, K. and {Bradley}, L.~J. and {Chaudry}, R.~A. and {Rees}, K. and {Tandy}, J.~A. and {Thomas}, P. and {Whillock}, M.~C.~R. and {Winter}, B. and {Doschek}, G.~A. and {Korendyke}, C.~M. and {Brown}, C.~M. and {Myers}, S. and {Mariska}, J. and {Seely}, J. and {Lang}, J. and {Kent}, B.~J. and {Shaughnessy}, B.~M. and {Young}, P.~R. and {Simnett}, G.~M. and {Castelli}, C.~M. and {Mahmoud}, S. and {Mapson-Menard}, H. and {Probyn}, B.~J. and {Thomas}, R.~J. and {Davila}, J. and {Dere}, K. and {Windt}, D. and {Shea}, J. and {Hagood}, R. and {Moye}, R. and {Hara}, H. and {Watanabe}, T. and {Matsuzaki}, K. and {Kosugi}, T. and {Hansteen}, V. and {Wikstol}, {\O}.},
        title = "{The EUV Imaging Spectrometer for Hinode}",
      journal = {\solphys},
         year = 2007,
        month = jun,
       volume = {243},
       number = {1},
        pages = {19-61},
          doi = {10.1007/s01007-007-0293-1},
       adsurl = {https://ui.adsabs.harvard.edu/abs/2007SoPh..243...19C}
}

@ARTICLE{Dacie:2016,
       author = {{Dacie}, S. and {D{\'e}moulin}, P. and {van Driel-Gesztelyi}, L. and {Long}, D.~M. and {Baker}, D. and {Janvier}, M. and {Yardley}, S.~L. and {P{\'e}rez-Su{\'a}rez}, D.},
        title = "{Evolution of the magnetic field distribution of active regions}",
      journal = {\aap},
         year = 2016,
        month = dec,
       volume = {596},
          eid = {A69},
        pages = {A69},
          doi = {10.1051/0004-6361/201628948},
archivePrefix = {arXiv},
       eprint = {1609.03723},
 primaryClass = {astro-ph.SR},
       adsurl = {https://ui.adsabs.harvard.edu/abs/2016A&A...596A..69D}
}

@ARTICLE{dejager:1986,
       author = {{de Jager}, O.~C. and {Raubenheimer}, B.~C. and {Swanepoel}, J.~W.~H.},
        title = "{Kernel density estimations applied to gamma ray light curves}",
      journal = {\aap},
         year = 1986,
        month = dec,
       volume = {170},
       number = {1},
        pages = {187-196},
       adsurl = {https://ui.adsabs.harvard.edu/abs/1986A&A...170..187D}
}

@ARTICLE{delzanna:2014,
       author = {{Del Zanna}, G. and {Mason}, H.~E.},
        title = "{Elemental abundances and temperatures of quiescent solar active region cores from X-ray observations}",
      journal = {\aap},
         year = 2014,
        month = may,
       volume = {565},
          eid = {A14},
        pages = {A14},
          doi = {10.1051/0004-6361/201423471},
       adsurl = {https://ui.adsabs.harvard.edu/abs/2014A&A...565A..14D}
}

@ARTICLE{delzanna:2021,
       author = {{Del Zanna}, G. and {Dere}, K.~P. and {Young}, P.~R. and {Landi}, E.},
        title = "{CHIANTI{\textemdash}An Atomic Database for Emission Lines. XVI. Version 10, Further Extensions}",
      journal = {\apj},
         year = 2021,
        month = mar,
       volume = {909},
       number = {1},
          eid = {38},
        pages = {38},
          doi = {10.3847/1538-4357/abd8ce},
archivePrefix = {arXiv},
       eprint = {2011.05211},
 primaryClass = {physics.atom-ph},
       adsurl = {https://ui.adsabs.harvard.edu/abs/2021ApJ...909...38D}
}

@ARTICLE{Dere:1997,
       author = {{Dere}, K.~P. and {Landi}, E. and {Mason}, H.~E. and {Monsignori Fossi}, B.~C. and {Young}, P.~R.},
        title = "{CHIANTI - an atomic database for emission lines}",
      journal = {\aaps},
         year = 1997,
        month = oct,
       volume = {125},
        pages = {149-173},
          doi = {10.1051/aas:1997368},
       adsurl = {https://ui.adsabs.harvard.edu/abs/1997A&AS..125..149D}
}

@ARTICLE{Doschek:2015,
       author = {{Doschek}, G.~A. and {Warren}, H.~P. and {Feldman}, U.},
        title = "{Anomalous Relative Ar/Ca Coronal Abundances Observed by the Hinode/EUV Imaging Spectrometer Near Sunspots}",
      journal = {\apjl},
         year = 2015,
        month = jul,
       volume = {808},
       number = {1},
          eid = {L7},
        pages = {L7},
          doi = {10.1088/2041-8205/808/1/L7},
       adsurl = {https://ui.adsabs.harvard.edu/abs/2015ApJ...808L...7D}
}

@ARTICLE{Doschek:2016,
       author = {{Doschek}, G.~A. and {Warren}, H.~P.},
        title = "{The Mysterious Case of the Solar Argon Abundance near Sunspots in Flares}",
      journal = {\apj},
         year = 2016,
        month = jul,
       volume = {825},
       number = {1},
          eid = {36},
        pages = {36},
          doi = {10.3847/0004-637X/825/1/36},
       adsurl = {https://ui.adsabs.harvard.edu/abs/2016ApJ...825...36D}
}

@ARTICLE{Doschek:2017,
       author = {{Doschek}, G.~A. and {Warren}, H.~P.},
        title = "{Sunspots, Starspots, and Elemental Abundances}",
      journal = {\apj},
         year = 2017,
        month = jul,
       volume = {844},
       number = {1},
          eid = {52},
        pages = {52},
          doi = {10.3847/1538-4357/aa7bea},
       adsurl = {https://ui.adsabs.harvard.edu/abs/2017ApJ...844...52D}
}

@ARTICLE{Feldman:1993,
       author = {{Feldman}, U. and {Widing}, K.~G.},
        title = "{Elemental Abundances in the Upper Solar Atmosphere of Quiet and Coronal Hole Regions (T E 4.3 X 10 5 K)}",
      journal = {\apj},
         year = 1993,
        month = sep,
       volume = {414},
        pages = {381},
          doi = {10.1086/173084},
       adsurl = {https://ui.adsabs.harvard.edu/abs/1993ApJ...414..381F}
}

@ARTICLE{Feldman:2009,
       author = {{Feldman}, U. and {Warren}, H.~P. and {Brown}, C.~M. and {Doschek}, G.~A.},
        title = "{Can the Composition of the Solar Corona Be Derived from Hinode/Extreme-Ultraviolet Imaging Spectrometer Spectra?}",
      journal = {\apj},
         year = 2009,
        month = apr,
       volume = {695},
       number = {1},
        pages = {36-45},
          doi = {10.1088/0004-637X/695/1/36},
       adsurl = {https://ui.adsabs.harvard.edu/abs/2009ApJ...695...36F}
}

@Article{Harris:2020,
 title         = {Array programming with {NumPy}},
 author        = {Charles R. Harris and K. Jarrod Millman and St{\'{e}}fan J.
                 van der Walt and Ralf Gommers and Pauli Virtanen and David
                 Cournapeau and Eric Wieser and Julian Taylor and Sebastian
                 Berg and Nathaniel J. Smith and Robert Kern and Matti Picus
                 and Stephan Hoyer and Marten H. van Kerkwijk and Matthew
                 Brett and Allan Haldane and Jaime Fern{\'{a}}ndez del
                 R{\'{i}}o and Mark Wiebe and Pearu Peterson and Pierre
                 G{\'{e}}rard-Marchant and Kevin Sheppard and Tyler Reddy and
                 Warren Weckesser and Hameer Abbasi and Christoph Gohlke and
                 Travis E. Oliphant},
 year          = {2020},
 month         = sep,
 journal       = {Nature},
 volume        = {585},
 number        = {7825},
 pages         = {357--362},
 doi           = {10.1038/s41586-020-2649-2},
 publisher     = {Springer Science and Business Media {LLC}},
 url           = {https://doi.org/10.1038/s41586-020-2649-2}
}

@Article{Hunter:2007,
  Author    = {Hunter, J. D.},
  Title     = {Matplotlib: A 2D graphics environment},
  Journal   = {Computing in Science \& Engineering},
  Volume    = {9},
  Number    = {3},
  Pages     = {90--95},
  publisher = {IEEE COMPUTER SOC},
  doi       = {10.1109/MCSE.2007.55},
  year      = 2007
}

@ARTICLE{Ko:2016,
       author = {{Ko}, Yuan-Kuen and {Young}, Peter R. and {Muglach}, Karin and {Warren}, Harry P. and {Ugarte-Urra}, Ignacio},
        title = "{Correlation of Coronal Plasma Properties and Solar Magnetic Field in a Decaying Active Region}",
      journal = {\apj},
         year = 2016,
        month = aug,
       volume = {826},
       number = {2},
          eid = {126},
        pages = {126},
          doi = {10.3847/0004-637X/826/2/126},
       adsurl = {https://ui.adsabs.harvard.edu/abs/2016ApJ...826..126K}
}

@ARTICLE{Kosugi:2007,
       author = {{Kosugi}, T. and {Matsuzaki}, K. and {Sakao}, T. and {Shimizu}, T. and {Sone}, Y. and {Tachikawa}, S. and {Hashimoto}, T. and {Minesugi}, K. and {Ohnishi}, A. and {Yamada}, T. and {Tsuneta}, S. and {Hara}, H. and {Ichimoto}, K. and {Suematsu}, Y. and {Shimojo}, M. and {Watanabe}, T. and {Shimada}, S. and {Davis}, J.~M. and {Hill}, L.~D. and {Owens}, J.~K. and {Title}, A.~M. and {Culhane}, J.~L. and {Harra}, L.~K. and {Doschek}, G.~A. and {Golub}, L.},
        title = "{The Hinode (Solar-B) Mission: An Overview}",
      journal = {\solphys},
         year = 2007,
        month = jun,
       volume = {243},
       number = {1},
        pages = {3-17},
          doi = {10.1007/s11207-007-9014-6},
       adsurl = {https://ui.adsabs.harvard.edu/abs/2007SoPh..243....3K}
}

@ARTICLE{Laming:2015,
       author = {{Laming}, J. Martin},
        title = "{The FIP and Inverse FIP Effects in Solar and Stellar Coronae}",
      journal = {Living Reviews in Solar Physics},
         year = 2015,
        month = dec,
       volume = {12},
       number = {1},
          eid = {2},
        pages = {2},
          doi = {10.1007/lrsp-2015-2},
archivePrefix = {arXiv},
       eprint = {1504.08325},
 primaryClass = {astro-ph.SR},
       adsurl = {https://ui.adsabs.harvard.edu/abs/2015LRSP...12....2L}
}

@ARTICLE{Long:2024,
       author = {{Long}, David M. and {Baker}, Deborah and {To}, Andy S.~H. and {van Driel-Gesztelyi}, Lidia and {Brooks}, David H. and {Stangalini}, Marco and {Murabito}, Mariarita and {James}, Alexander W. and {Mathioudakis}, Mihalis and {Testa}, Paola},
        title = "{Identifying Plasma Fractionation Processes in the Chromosphere Using IRIS}",
      journal = {\apj},
         year = 2024,
        month = apr,
       volume = {965},
       number = {1},
          eid = {63},
        pages = {63},
          doi = {10.3847/1538-4357/ad3234},
archivePrefix = {arXiv},
       eprint = {2403.06711},
 primaryClass = {astro-ph.SR},
       adsurl = {https://ui.adsabs.harvard.edu/abs/2024ApJ...965...63L}
}

@ARTICLE{Meyer:1985,
       author = {{Meyer}, J. -P.},
        title = "{The baseline composition of solar energetic particles}",
      journal = {\apjs},
         year = 1985,
        month = jan,
       volume = {57},
        pages = {151-171},
          doi = {10.1086/191000},
       adsurl = {https://ui.adsabs.harvard.edu/abs/1985ApJS...57..151M}
}

@ARTICLE{Mihailescu:2022,
       author = {{Mihailescu}, Teodora and {Baker}, Deborah and {Green}, Lucie M. and {van Driel-Gesztelyi}, Lidia and {Long}, David M. and {Brooks}, David H. and {To}, Andy S.~H.},
        title = "{What Determines Active Region Coronal Plasma Composition?}",
      journal = {\apj},
         year = 2022,
        month = jul,
       volume = {933},
       number = {2},
          eid = {245},
        pages = {245},
          doi = {10.3847/1538-4357/ac6e40},
archivePrefix = {arXiv},
       eprint = {2205.05027},
 primaryClass = {astro-ph.SR},
       adsurl = {https://ui.adsabs.harvard.edu/abs/2022ApJ...933..245M}
}

@ARTICLE{Mueller:2020,
       author = {{M{\"u}ller}, D. and {St. Cyr}, O.~C. and {Zouganelis}, I. and {Gilbert}, H.~R. and {Marsden}, R. and {Nieves-Chinchilla}, T. and {Antonucci}, E. and {Auch{\`e}re}, F. and {Berghmans}, D. and {Horbury}, T.~S. and {Howard}, R.~A. and {Krucker}, S. and {Maksimovic}, M. and {Owen}, C.~J. and {Rochus}, P. and {Rodriguez-Pacheco}, J. and {Romoli}, M. and {Solanki}, S.~K. and {Bruno}, R. and {Carlsson}, M. and {Fludra}, A. and {Harra}, L. and {Hassler}, D.~M. and {Livi}, S. and {Louarn}, P. and {Peter}, H. and {Sch{\"u}hle}, U. and {Teriaca}, L. and {del Toro Iniesta}, J.~C. and {Wimmer-Schweingruber}, R.~F. and {Marsch}, E. and {Velli}, M. and {De Groof}, A. and {Walsh}, A. and {Williams}, D.},
        title = "{The Solar Orbiter mission. Science overview}",
      journal = {\aap},
         year = 2020,
        month = oct,
       volume = {642},
          eid = {A1},
        pages = {A1},
          doi = {10.1051/0004-6361/202038467},
archivePrefix = {arXiv},
       eprint = {2009.00861},
 primaryClass = {astro-ph.SR},
       adsurl = {https://ui.adsabs.harvard.edu/abs/2020A&A...642A...1M}
}

@ARTICLE{Murabito:2021,
       author = {{Murabito}, M. and {Stangalini}, M. and {Baker}, D. and {Valori}, G. and {Jess}, D.~B. and {Jafarzadeh}, S. and {Brooks}, D.~H. and {Ermolli}, I. and {Giorgi}, F. and {Grant}, S.~D.~T. and {Long}, D.~M. and {van Driel-Gesztelyi}, L.},
        title = "{Investigating the origin of magnetic perturbations associated with the FIP Effect}",
      journal = {\aap},
         year = 2021,
        month = dec,
       volume = {656},
          eid = {A87},
        pages = {A87},
          doi = {10.1051/0004-6361/202141504},
archivePrefix = {arXiv},
       eprint = {2108.11164},
 primaryClass = {astro-ph.SR},
       adsurl = {https://ui.adsabs.harvard.edu/abs/2021A&A...656A..87M}
}

@ARTICLE{Murabito:2024,
       author = {{Murabito}, Mariarita and {Stangalini}, Marco and {Laming}, J. Martin and {Baker}, Deborah and {To}, Andy S.~H. and {Long}, David M. and {Brooks}, David H. and {Jafarzadeh}, Shahin and {Jess}, David B. and {Valori}, Gherardo},
        title = "{Observation of Alfv{\'e}n Wave Reflection in the Solar Chromosphere: Ponderomotive Force and First Ionization Potential Effect}",
      journal = {\prl},
         year = 2024,
        month = may,
       volume = {132},
       number = {21},
          eid = {215201},
        pages = {215201},
          doi = {10.1103/PhysRevLett.132.215201},
archivePrefix = {arXiv},
       eprint = {2404.08305},
 primaryClass = {astro-ph.SR},
       adsurl = {https://ui.adsabs.harvard.edu/abs/2024PhRvL.132u5201M}
}

@ARTICLE{Orlovskij:2025,
       author = {{Orlovskij}, Dominik and {To}, Andy S.~H. and {Long}, David M.},
        title = "{Signs of Sulphur fractionation under high magnetic field strength}",
      journal = {\ptrsa},
         year = 2025,

}

@ARTICLE{Owen:2020,
       author = {{Owen}, C.~J. and {Bruno}, R. and {Livi}, S. and {Louarn}, P. and {Al Janabi}, K. and {Allegrini}, F. and {Amoros}, C. and {Baruah}, R. and {Barthe}, A. and {Berthomier}, M. and {Bordon}, S. and {Brockley-Blatt}, C. and {Brysbaert}, C. and {Capuano}, G. and {Collier}, M. and {DeMarco}, R. and {Fedorov}, A. and {Ford}, J. and {Fortunato}, V. and {Fratter}, I. and {Galvin}, A.~B. and {Hancock}, B. and {Heirtzler}, D. and {Kataria}, D. and {Kistler}, L. and {Lepri}, S.~T. and {Lewis}, G. and {Loeffler}, C. and {Marty}, W. and {Mathon}, R. and {Mayall}, A. and {Mele}, G. and {Ogasawara}, K. and {Orlandi}, M. and {Pacros}, A. and {Penou}, E. and {Persyn}, S. and {Petiot}, M. and {Phillips}, M. and {P{\v{r}}ech}, L. and {Raines}, J.~M. and {Reden}, M. and {Rouillard}, A.~P. and {Rousseau}, A. and {Rubiella}, J. and {Seran}, H. and {Spencer}, A. and {Thomas}, J.~W. and {Trevino}, J. and {Verscharen}, D. and {Wurz}, P. and {Alapide}, A. and {Amoruso}, L. and {Andr{\'e}}, N. and {Anekallu}, C. and {Arciuli}, V. and {Arnett}, K.~L. and {Ascolese}, R. and {Bancroft}, C. and {Bland}, P. and {Brysch}, M. and {Calvanese}, R. and {Castronuovo}, M. and {{\v{C}}erm{\'a}k}, I. and {Chornay}, D. and {Clemens}, S. and {Coker}, J. and {Collinson}, G. and {D'Amicis}, R. and {Dandouras}, I. and {Darnley}, R. and {Davies}, D. and {Davison}, G. and {De Los Santos}, A. and {Devoto}, P. and {Dirks}, G. and {Edlund}, E. and {Fazakerley}, A. and {Ferris}, M. and {Frost}, C. and {Fruit}, G. and {Garat}, C. and {G{\'e}not}, V. and {Gibson}, W. and {Gilbert}, J.~A. and {de Giosa}, V. and {Gradone}, S. and {Hailey}, M. and {Horbury}, T.~S. and {Hunt}, T. and {Jacquey}, C. and {Johnson}, M. and {Lavraud}, B. and {Lawrenson}, A. and {Leblanc}, F. and {Lockhart}, W. and {Maksimovic}, M. and {Malpus}, A. and {Marcucci}, F. and {Mazelle}, C. and {Monti}, F. and {Myers}, S. and {Nguyen}, T. and {Rodriguez-Pacheco}, J. and {Phillips}, I. and {Popecki}, M. and {Rees}, K. and {Rogacki}, S.~A. and {Ruane}, K. and {Rust}, D. and {Salatti}, M. and {Sauvaud}, J.~A. and {Stakhiv}, M.~O. and {Stange}, J. and {Stubbs}, T. and {Taylor}, T. and {Techer}, J. -D. and {Terrier}, G. and {Thibodeaux}, R. and {Urdiales}, C. and {Varsani}, A. and {Walsh}, A.~P. and {Watson}, G. and {Wheeler}, P. and {Willis}, G. and {Wimmer-Schweingruber}, R.~F. and {Winter}, B. and {Yardley}, J. and {Zouganelis}, I.},
        title = "{The Solar Orbiter Solar Wind Analyser (SWA) suite}",
      journal = {\aap},
         year = 2020,
        month = oct,
       volume = {642},
          eid = {A16},
        pages = {A16},
          doi = {10.1051/0004-6361/201937259},
       adsurl = {https://ui.adsabs.harvard.edu/abs/2020A&A...642A..16O}
}

@ARTICLE{Reep:2025,
       author = {{Reep}, Jeffrey W. and {Benavitz}, Luke F. and {To}, Andy S.~H. and {Brooks}, David H. and {Laming}, J.~Martin and {Antolin}, Patrick and {Long}, David M. and {Baker}, Deborah and {Tarr}, Lucas A.},
        title = "{Radiative hydrodynamic simulations of FIP fractionation in solar flares}",
      journal = {\ptrsa},
         year = 2025,

}

@ARTICLE{Scott:2015a,
       author = {{Scott}, Pat and {Grevesse}, Nicolas and {Asplund}, Martin and {Sauval}, A. Jacques and {Lind}, Karin and {Takeda}, Yoichi and {Collet}, Remo and {Trampedach}, Regner and {Hayek}, Wolfgang},
        title = "{The elemental composition of the Sun. I. The intermediate mass elements Na to Ca}",
      journal = {\aap},
         year = 2015,
        month = jan,
       volume = {573},
          eid = {A25},
        pages = {A25},
          doi = {10.1051/0004-6361/201424109},
archivePrefix = {arXiv},
       eprint = {1405.0279},
 primaryClass = {astro-ph.SR},
       adsurl = {https://ui.adsabs.harvard.edu/abs/2015A&A...573A..25S}
}

@ARTICLE{Scott:2015b,
       author = {{Scott}, Pat and {Asplund}, Martin and {Grevesse}, Nicolas and {Bergemann}, Maria and {Sauval}, A. Jacques},
        title = "{The elemental composition of the Sun. II. The iron group elements Sc to Ni}",
      journal = {\aap},
         year = 2015,
        month = jan,
       volume = {573},
          eid = {A26},
        pages = {A26},
          doi = {10.1051/0004-6361/201424110},
archivePrefix = {arXiv},
       eprint = {1405.0287},
 primaryClass = {astro-ph.SR},
       adsurl = {https://ui.adsabs.harvard.edu/abs/2015A&A...573A..26S}
}

@BOOK{Silverman:1986,
       author = {{Silverman}, B.~W.},
        title = "{Density estimation for statistics and data analysis}",
         year = 1986,
       adsurl = {https://ui.adsabs.harvard.edu/abs/1986desd.book.....S}
}

@ARTICLE{Spice:2020,
       author = {{SPICE Consortium} and {Anderson}, M. and {Appourchaux}, T. and {Auch{\`e}re}, F. and {Aznar Cuadrado}, R. and {Barbay}, J. and {Baudin}, F. and {Beardsley}, S. and {Bocchialini}, K. and {Borgo}, B. and {Bruzzi}, D. and {Buchlin}, E. and {Burton}, G. and {B{\"u}chel}, V. and {Caldwell}, M. and {Caminade}, S. and {Carlsson}, M. and {Curdt}, W. and {Davenne}, J. and {Davila}, J. and {Deforest}, C.~E. and {Del Zanna}, G. and {Drummond}, D. and {Dubau}, J. and {Dumesnil}, C. and {Dunn}, G. and {Eccleston}, P. and {Fludra}, A. and {Fredvik}, T. and {Gabriel}, A. and {Giunta}, A. and {Gottwald}, A. and {Griffin}, D. and {Grundy}, T. and {Guest}, S. and {Gyo}, M. and {Haberreiter}, M. and {Hansteen}, V. and {Harrison}, R. and {Hassler}, D.~M. and {Haugan}, S.~V.~H. and {Howe}, C. and {Janvier}, M. and {Klein}, R. and {Koller}, S. and {Kucera}, T.~A. and {Kouliche}, D. and {Marsch}, E. and {Marshall}, A. and {Marshall}, G. and {Matthews}, S.~A. and {McQuirk}, C. and {Meining}, S. and {Mercier}, C. and {Morris}, N. and {Morse}, T. and {Munro}, G. and {Parenti}, S. and {Pastor-Santos}, C. and {Peter}, H. and {Pfiffner}, D. and {Phelan}, P. and {Philippon}, A. and {Richards}, A. and {Rogers}, K. and {Sawyer}, C. and {Schlatter}, P. and {Schmutz}, W. and {Sch{\"u}hle}, U. and {Shaughnessy}, B. and {Sidher}, S. and {Solanki}, S.~K. and {Speight}, R. and {Spescha}, M. and {Szwec}, N. and {Tamiatto}, C. and {Teriaca}, L. and {Thompson}, W. and {Tosh}, I. and {Tustain}, S. and {Vial}, J. -C. and {Walls}, B. and {Waltham}, N. and {Wimmer-Schweingruber}, R. and {Woodward}, S. and {Young}, P. and {de Groof}, A. and {Pacros}, A. and {Williams}, D. and {M{\"u}ller}, D.},
        title = "{The Solar Orbiter SPICE instrument. An extreme UV imaging spectrometer}",
      journal = {\aap},
         year = 2020,
        month = oct,
       volume = {642},
          eid = {A14},
        pages = {A14},
          doi = {10.1051/0004-6361/201935574},
archivePrefix = {arXiv},
       eprint = {1909.01183},
 primaryClass = {astro-ph.IM},
       adsurl = {https://ui.adsabs.harvard.edu/abs/2020A&A...642A..14S}
}

@ARTICLE{Stangalini:2021,
       author = {{Stangalini}, M. and {Baker}, D. and {Valori}, G. and {Jess}, D.~B. and {Jafarzadeh}, S. and {Murabito}, M. and {To}, A.~S.~H. and {Brooks}, D.~H. and {Ermolli}, I. and {Giorgi}, F. and {MacBride}, C.~D.},
        title = "{Spectropolarimetric fluctuations in a sunspot chromosphere}",
      journal = {Philosophical Transactions of the Royal Society of London Series A},
         year = 2021,
        month = feb,
       volume = {379},
       number = {2190},
          eid = {20200216},
        pages = {20200216},
          doi = {10.1098/rsta.2020.0216},
archivePrefix = {arXiv},
       eprint = {2009.05302},
 primaryClass = {astro-ph.SR},
       adsurl = {https://ui.adsabs.harvard.edu/abs/2021RSPTA.37900216S}
}

@ARTICLE{Sunpy:2020,
  doi = {10.3847/1538-4357/ab4f7a},
  url = {https://iopscience.iop.org/article/10.3847/1538-4357/ab4f7a},
  author = {{The SunPy Community} and Barnes, Will T. and Bobra, Monica G. and Christe, Steven D. and Freij, Nabil and Hayes, Laura A. and Ireland, Jack and Mumford, Stuart and Perez-Suarez, David and Ryan, Daniel F. and Shih, Albert Y. and Chanda, Prateek and Glogowski, Kolja and Hewett, Russell and Hughitt, V. Keith and Hill, Andrew and Hiware, Kaustubh and Inglis, Andrew and Kirk, Michael S. F. and Konge, Sudarshan and Mason, James Paul and Maloney, Shane Anthony and Murray, Sophie A. and Panda, Asish and Park, Jongyeob and Pereira, Tiago M. D. and Reardon, Kevin and Savage, Sabrina and Sipőcz, Brigitta M. and Stansby, David and Jain, Yash and Taylor, Garrison and Yadav, Tannmay and Rajul and Dang, Trung Kien},
  title = {The SunPy Project: Open Source Development and Status of the Version 1.0 Core Package},
  journal = {The Astrophysical Journal},
  volume = {890},
  issue = {1},
  pages = {68-},
  publisher = {American Astronomical Society},
  year = {2020}
}

@ARTICLE{Testa:2023,
       author = {{Testa}, Paola and {Mart{\'\i}nez-Sykora}, Juan and {De Pontieu}, Bart},
        title = "{Coronal Abundances in an Active Region: Evolution and Underlying Chromospheric and Transition Region Properties}",
      journal = {\apj},
         year = 2023,
        month = feb,
       volume = {944},
       number = {2},
          eid = {117},
        pages = {117},
          doi = {10.3847/1538-4357/acb343},
archivePrefix = {arXiv},
       eprint = {2211.07755},
 primaryClass = {astro-ph.SR},
       adsurl = {https://ui.adsabs.harvard.edu/abs/2023ApJ...944..117T}
}

@ARTICLE{To:2021,
       author = {{To}, Andy S.~H. and {Long}, David M. and {Baker}, Deborah and {Brooks}, David H. and {van Driel-Gesztelyi}, Lidia and {Laming}, J. Martin and {Valori}, Gherardo},
        title = "{The Evolution of Plasma Composition during a Solar Flare}",
      journal = {\apj},
         year = 2021,
        month = apr,
       volume = {911},
       number = {2},
          eid = {86},
        pages = {86},
          doi = {10.3847/1538-4357/abe85a},
archivePrefix = {arXiv},
       eprint = {2102.09985},
 primaryClass = {astro-ph.SR},
       adsurl = {https://ui.adsabs.harvard.edu/abs/2021ApJ...911...86T}
}

@ARTICLE{To:2023,
       author = {{To}, Andy S.~H. and {James}, Alexander W. and {Bastian}, T.~S. and {van Driel-Gesztelyi}, Lidia and {Long}, David M. and {Baker}, Deborah and {Brooks}, David H. and {Lomuscio}, Samantha and {Stansby}, David and {Valori}, Gherardo},
        title = "{Understanding the Relationship between Solar Coronal Abundances and F10.7 cm Radio Emission}",
      journal = {\apj},
         year = 2023,
        month = may,
       volume = {948},
       number = {2},
          eid = {121},
        pages = {121},
          doi = {10.3847/1538-4357/acbc1b},
archivePrefix = {arXiv},
       eprint = {2304.02552},
 primaryClass = {astro-ph.SR},
       adsurl = {https://ui.adsabs.harvard.edu/abs/2023ApJ...948..121T}
}

@ARTICLE{To:2024,
       author = {{To}, Andy S.~H. and {Brooks}, David H. and {Imada}, Shinsuke and {French}, Ryan J. and {van Driel-Gesztelyi}, Lidia and {Baker}, Deborah and {Long}, David M. and {Ashfield}, IV, William and {Hayes}, Laura A.},
        title = "{Spatially resolved plasma composition evolution in a solar flare {\textendash} The effect of reconnection outflow}",
      journal = {\aap},
         year = 2024,
        month = nov,
       volume = {691},
          eid = {A95},
        pages = {A95},
          doi = {10.1051/0004-6361/202449246},
archivePrefix = {arXiv},
       eprint = {2409.18188},
 primaryClass = {astro-ph.SR},
       adsurl = {https://ui.adsabs.harvard.edu/abs/2024A&A...691A..95T}
}

\end{document}